\documentclass[prd,aps,floatfix,nofootinbib,twocolumn,tightenlines,showpacs]{revtex4}
\usepackage{dcolumn}
\usepackage{latexsym}
\usepackage{graphicx}
\usepackage{bm}

\def\G{\Gamma}

\def\co{{\cal O}}

\def\svev#1{\left\langle #1\right\rangle}       
\def\tr{{\rm tr}\,}

\long \def \blockcomment #1\endcomment{}

\begin{document}

\title{%
Zero of the discrete beta function in SU(3) lattice gauge theory with color sextet fermions
}
\author{Yigal Shamir}
\author{Benjamin Svetitsky}
 \affiliation{Raymond and Beverly Sackler School of Physics and Astronomy,  Tel~Aviv University, 69978
Tel~Aviv, Israel}

\author{Thomas DeGrand}%
 \affiliation{Department of Physics,
University of Colorado, Boulder, CO 80309, USA}

\begin{abstract}
We have carried out a Schrodinger functional (SF) calculation for the
SU(3) lattice gauge theory with two flavors of Wilson fermions in the sextet representation of
 the gauge group.
  We find that the discrete beta function, which governs the change in the running coupling
under a discrete change of spatial scale, changes sign when the SF renormalized coupling is in
 the neighborhood of $g^2=2.0$. 
 The simplest explanation is that the theory has an
 infrared-attractive fixed point, but 
more complicated possibilities are allowed by the data.
While we compare rescalings by factors of 2 and $4/3$,
we work at a single lattice spacing.
\end{abstract}

\maketitle

\section{Introduction}
Gauge theories with groups larger than SU(3), or with light fermions in representations higher than the
fundamental, are a staple of theories that go beyond the Standard Model \cite{Hill:2002ap}.
Among the mechanisms proposed to connect these theories to the Standard Model at low energies
are technicolor \cite{Weinberg:1975gm,Susskind:1978ms} and tumbling \cite{Raby:1979my}, 
with many associated variants.  Both of these depend largely on weak-coupling pictures for
their dynamics: a perturbative $\beta$ function to take technicolor from weak to strong 
coupling as the energy scale drops, and a most-attractive-channel argument for scale separation
and selection of the condensed channel in tumbling.
Nonperturbative tests of these pictures are long overdue.
While some lattice studies have been carried out on SU(3) gauge theories with $N_f>3$ 
fundamental flavors \cite{Brown:1992fz, Damgaard:1997ut, Iwasaki:2003de, Appelquist:2007hu,Deuzeman:2008sc},
scant attention has been paid to more general alternatives where a richer set of phenomena may be
sought~\cite{Kogut:1984sb, Kogut:1985xa, Karsch:1998qj,Catterall:2007yx}. 

A number of ideas focus on the behavior of the gauge theory's $\beta$ function and the possibility of an IR-attractive fixed point~\cite{Caswell:1974gg,Banks:1981nn}. 
The infrared limit of the massless theory is then
scale-invariant and probably conformal, devoid of confinement and 
of chiral symmetry breaking~\cite{Appelquist:1998xf,Appelquist:1998rb}.
Alternatively, the
 near-appearance of a fixed point makes the $\beta$ function hover near the axis without crossing it; this is the 
scenario of ``walking''~\cite{Bando:1987we,Cohen:1988sq}, wherein enormous scale ratios are generated before confinement finally sets in at large distances.

Let us describe the perturbative picture of the theory we have 
studied.  This is the SU(3) gauge theory with $N_f$ fermions in the symmetric two-index 
representation \cite{Sannino:2004qp, Dietrich:2006cm}, which for this group is the sextet.
Consider varying $N_f$ upwards from zero.
At first, the one- and two-loop terms in the $\beta$ function~\cite{Caswell:1974gg,Jones:1974mm},
\begin{equation}
\beta(g^2)=\frac{dg^2}{d\log(q^2)}=-\frac{b_1}{16\pi^2}g^4-\frac{b_2}{(16\pi^2)^2}g^6+\cdots,
\label{2loopbeta}
\end{equation}
where
\begin{eqnarray}
b_1&=&11-\frac{10}3N_f\\
b_2&=&102-\frac{250}3N_f,
\end{eqnarray}
are both negative: this is like ordinary QCD.
When $N_f$ passes $\frac{306}{250}\simeq1.22$, the two-loop coefficient in
 the $\beta$ function becomes positive and hence the two-loop $\beta$ function 
acquires a zero at positive coupling $g=g^*$.  
For the $N_f=2$ theory studied in this paper, this  zero is at the very large coupling $g^2\simeq10.4$. 
As $N_f$ grows $g^*$ becomes weaker,  
lending perhaps more credibility
to the perturbative prediction \cite{Appelquist:1988yc,Appelquist:1996dq}. 
It should be kept in mind, though, that when one-loop
and two-loop effects compete, there are usually similar-size contributions coming from higher orders in
perturbation theory. An infrared fixed point generally requires nonperturbative confirmation.

We have begun a study of the SU(3) lattice theory with $N_f=2$ Wilson fermions in the sextet representation.
Since the $\beta$ function gives the most direct approach to all these scenarios,
we chose it as our first object of study. We apply the
 Schr\"odinger functional (SF) method 
\cite{Luscher:1992an,Luscher:1993gh, Sint:1993un, Sint:1995ch, 
Luscher:1996sc, Sommer:1997xw, Jansen:1998mx, DellaMorte:2004bc},
 wherein we impose a background gauge field and calculate directly the running coupling 
as the scale of the background field is changed.  Since we change the scale by a discrete
 amount, we obtain the discrete beta function (DBF), analogous to the usual $\beta$ function.
 
We find that the DBF of the massless theory crosses zero at $g^2\simeq2.0$,
far short of the perturbative prediction.  
  If the full renormalization-group flow of the theory can indeed be summed up by this single
 coupling constant, then this is an IR-attractive fixed point, implying scale invariance in
 the IR physics of the strictly massless lattice theory defined in the fixed point's catchment basin.  
 At the same time, the zero of the DBF allows for more complex possibilities stemming from more complex RG flows, as discussed below.  
 
In order to judge the significance of the zero of the DBF, as well as to study the physics of the theory in its vicinity, we have also calculated observables connected with the $q\bar q$ potential and with chiral symmetry.  
While our presentation here is brief, we wish to place it in its proper context.
A simple and general argument shows that the formation of any bound states made out of light fermions in a gauge theory (without scalars)
necessitates the spontaneous breaking of chiral symmetry \cite{Casher:1979vw}.
Thus, if the infrared limit is a confining theory, chiral symmetry must be broken
spontaneously at an energy scale at least as high as the confinement 
scale~\cite{Casher:1979vw,Banks:1979yr}.%
\footnote{Casimir scaling is
the most popular mechanism to explain how chiral symmetry could 
be broken at an energy scale much higher than
the confinement scale~\cite{Kogut:1983sm,Kogut:1984nq,Kogut:1984sb,Kogut:1985xa,Karsch:1998qj}.}
If the two scales are indeed separated, the intermediate region breaks chiral symmetry but does not show confinement; and an entirely nonconfining theory can still break chiral symmetry spontaneously.
A conformal theory, on the other hand, is inconsistent with spontaneously broken chiral symmetry since the breaking creates a mass scale.

\section{Schrodinger functional method}

In order to introduce the SF running coupling $g(L)$, we begin with the gauge theory
 defined in a small Euclidean box of volume $L^4$.
We can consistently choose the coupling in this volume to be small;
asymptotic freedom ensures that there is only one effective coupling, that it runs
 with the perturbative $\beta$ function, and hence that at yet smaller 
distances, the coupling is even smaller.
In determining the running coupling
{\em non\/}-perturbatively,  virtually any observable
can in principle be used to extract it.
For consistency, we require that a perturbative calculation of the same 
observable will indeed reproduce the running coupling at this small scale.

The SF definition of the running coupling $g(L)$ is an application
of the background field method.  Consider a background field calculation
in the classical field strength $F_{\mu\nu}/g$.
If by construction the only distance scale that characterizes the background field is $L$,
the $n$-loop effective action $\Gamma\equiv-\log Z$ gives the running coupling via
\begin{equation}
\label{Gamma}
\Gamma = g(L)^{-2} S_{YM}^{cl} , 
\end{equation}
where
\begin{equation}
\label{SYM}
S_{YM}^{cl} = \int d^4x\, F^2_{\mu\nu}\, .
\end{equation}
$g(L)$ is the result of integrating the $n$-loop $\beta$ function.

The SF defines the background field by imposing
Dirichlet boundary conditions at $t=0$ and $t=L$.
The boundary values are chosen
such that the classical action has a 
unique, nontrivial minimum, and the configuration at this minimum is $F_{\mu\nu}/g$ \cite{Luscher:1992an}.  The effective action $\Gamma$ is
then calculated and compared via Eq.~(\ref{Gamma}) 
to $S_{YM}^{cl}$, which in turn is to be evaluated for the classical
 field that minimizes it with the given boundary conditions.
 If the effective action is calculated nonperturbatively, this procedure gives a nonperturbative definition of $g(L)$.

At short distances, the static potential between color sources
is a Coulomb potential.  The scale dependence of $g^2$ provides a
small correction.  We postulate that this is true
when the volume is small enough that confinement is not evident.
In large volumes, on the other hand,
the static potential can be qualitatively different;
the notion of a unique effective coupling that depends only
on the overall scale may no longer be tenable.  The upshot is that one 
must be cautious in drawing conclusions from the running of a single coupling constant.  We bear this in mind as we take Eq.~(\ref{Gamma}) to define of $g(L)$ beyond perturbation theory.

The continuum framework carries over
to the lattice with minor adaptations.  A technical obstacle
is that Monte Carlo methods do not allow for the direct computation
of the effective action.  This is solved by considering
a family of gauge-field boundary values that depend on a continuous parameter
$\eta$.  
By differentiating Eq.~(\ref{Gamma}) we obtain
\begin{equation}
\label{defg}
\left.\frac{\partial \G}{\partial\eta} \right|_{\eta=0}
= \frac{K}{g^2(L)}\,,
\qquad
K \equiv \left.\frac{\partial S_{YM}^{cl}}{\partial\eta} \right|_{\eta=0} \,.
\end{equation}
The derivative of $\Gamma$ gives an observable quantity, while $K$ is just a number
\cite{Luscher:1993gh}.

Our lattice theory is defined by the single-plaquette gauge action and a 
Wilson fermion action with added clover term~\cite{Sheikholeslami:1985ij}.  
Our choice of $N_f=2$ ensures  positivity of the fermion determinant, which in turn allows use of
the standard hybrid Monte Carlo algorithm.
We eschew perturbative corrections  \cite{Luscher:1996sc} to the parameters and operators in this exploratory study, 
but we do modify the clover term's coefficient via 
self-consistent tadpole improvement, $c_{SW}=1/u_0^3$.
The quantum correction $c_{SW}-1$
is known to be nonperturbative and large for fundamental fermions 
\cite{Jansen:1998mx}, where it serves to bring the theory closer to the continuum limit;
the tadpole estimate supplies a large portion of the correction.

The SF boundary conditions take the form of fixed,
spatially constant values for the spacelike 
links $U_i$ on the top and bottom layers of the lattice.  These links enter 
into the gauge plaquette and into the clover term of the fermion action.
Thus the $\eta$ derivative of the effective action is given by
\begin{equation}
  \left.\frac{\partial \Gamma}{\partial\eta} \right|_{\eta=0}
  =
  \left.\svev{\frac{\partial S_{YM}}{\partial\eta}
  -\tr \left( \frac{1}{D_F^\dagger}\;
        \frac{\partial (D_F^\dagger D_F)}{\partial\eta}\;
            \frac{1}{D_F} \right)}\right|_{\eta=0},
            \label{deta}
\end{equation}
where $D_F$ is the complete Wilson-clover fermion action.
(The appearance of $D^\dagger_FD_F$ indicates that $N_f=2$; we evaluate
the functional trace with a noisy estimator \cite{DellaMorte:2004bc}.)
The boundary fields are chosen as described in Ref.~\cite{Jansen:1998mx};
the parameter $\eta$ enters linearly into phase angles so the derivatives 
on the right-hand side of Eq.~(\ref{deta}) are implemented by putting 
the appropriate fixed values in place of the boundary links \cite{Luscher:1993gh}.
With these boundary values the coefficient%
\footnote{This is the value for a lattice with $8^4$ sites.  There is a tiny
 dependence on the lattice size, which we safely neglect.} $K=37.7$.
We also impose twisted spatial boundary conditions on the fermion fields 
as suggested in Ref.~\cite{Sint:1995ch},
$\psi(x+L)=\exp(i\theta)\psi(x)$, with $\theta=\pi/5$ on all three 
axes \cite{DellaMorte:2004bc}.
 
Equation~(\ref{defg}) defines the running coupling at any given length 
scale $L$, which we take to be the linear size of the lattice.  
The {\em discrete beta function\/} gives the change in $K/g^2$ 
when $L$ is multiplied by $n$.  Defining $u\equiv K/g^2(L)$, we write
\begin{equation}
B(u,n)=\frac K{g^2(nL)}-u,
\label{Beta}
\end{equation}
which is the counterpart of Eq.~(\ref{2loopbeta}).%
\footnote{In the SF literature it is customary to define the {\em step scaling function\/} $\sigma(v)=g^2(2L)$, with $v\equiv g^2(L)$.}
  We use a lattice
 approximation of Eq.~(\ref{Beta}), which introduces an implicit dependence on the ultraviolet cutoff:
We calculate $B(u,n=2)$ with lattice spacing $a=L/4$, and $B(u,n=4/3)$ with $a=L/6$.

\section{Lattice calculation of the discrete beta function}

We study herein the scaling of only the massless theory. 
 In the case of Wilson fermions, where the quark mass is unprotected
 against additive renormalization,
 this means fixing the hopping parameter to its critical value, $\kappa=\kappa_c$,
 at each value of the bare lattice coupling $\beta\equiv6/g_0^2$.
It is easy to locate $\kappa_c(\beta)$ when SF boundary conditions are used, 
since these boundary conditions  (and the spatial twists) limit the condition
 number of the fermion matrix even when there is no mass, and hence we can 
carry out simulations precisely at $\kappa_c$.%
 A straightforward way of finding $\kappa_c$ is to calculate the quark
 mass as defined by the lattice approximation of an axial Ward identity (AWI),
\begin{equation}
m_q=\frac12\,\frac{\partial_4\svev{A^b_4(t)\co^b(0)}}{\svev{P^b(t)\co^b(0)}}.
\label{AWI}
\end{equation}
Here $A^b_4(t)=\bar\psi\gamma_5\gamma_4\tau^b\psi$ is the time component of the 
local axial vector current with flavor $b$, taken at zero spatial momentum on 
the time slice $t$; $P^b(t)$ is the local pseudoscalar density.
The operator $\co^b(0)$ is defined by introducing spatially constant (Grassmann)
Dirichlet boundary conditions for the fermions, differentiating
with respect to these boundary values, and finally setting them to zero
\cite{Luscher:1996sc}. 
The derivative in Eq.~(\ref{AWI}) is a symmetric difference evaluated 
about $t=L/2$, the center of the lattice.  Equation~(\ref{AWI}) neglects 
multiplicative renormalization of the currents, but this is unimportant 
since we only use it to locate $\kappa_c$ by demanding $m_q=0$.
In addition, we neglect the mixing of $A^b_\mu$ with $\partial_\mu P^b$,
which is known to be a small effect for fundamental
fermions \cite{Luscher:1996sc,Jansen:1998mx}.

We list the values of $\kappa_c$ and the tadpole coefficient
$u_0$ in Table~\ref{Table1}.  While these values were determined
 for $L=4a$, we find a small movement in $m_q$ 
(0.01--0.04) when going to $L=6a$ and then $L=8a$.
Calculation of the DBF demands keeping the ultraviolet cutoff fixed as the volume is changed, and thus $\beta$, $\kappa$, and~$u_0$ {\em must\/}
remain unchanged when comparing different lattices.
The undesirable shift in $m_q$ is a discretization error; it can only be reduced by going to larger pairs of lattices.

\begin{table}[t]
\caption{$\kappa_c$ and~$u_0$ for $L=4a$ with SF boundary conditions.  
Linear interpolation may be used safely between $\beta=5.0$ and~5.5 and between $\beta=5.5$ and~6.0.
\label{Table1}}
\begin{ruledtabular}
\begin{tabular}{ccc}
$\beta$ & $\ \kappa_c\ $ & $\ u_0\ $ \\
\hline
5.0		&.1723	&.875\\
5.5		&.1654	&.887\\
6.0		&.1610	&.900\\
7.0		&.1536	&.916\\
8.0		&.1486	&.928\\
\end{tabular}
\end{ruledtabular}
\end{table}

\begin{figure}[htb]
\vskip\baselineskip
\includegraphics*[width=\columnwidth]
{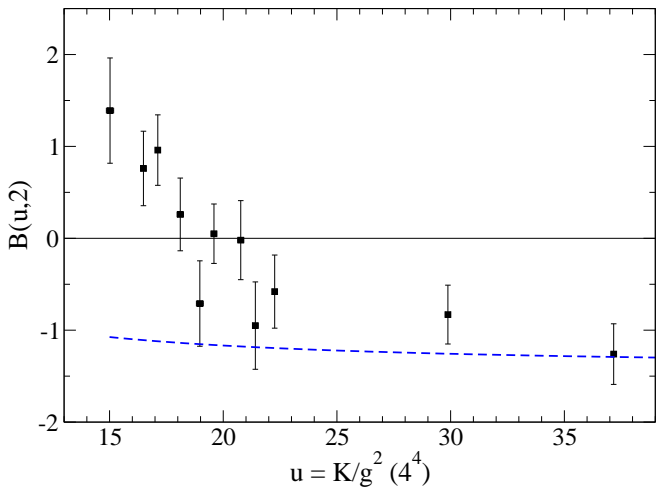}
\caption{Discrete beta function for the scale transformation $L\to2L$, as 
defined in Eq.~(\ref{Beta}).  The lattice spacing is fixed such that $L=4a$.  
The dashed curve is the two-loop result. The data points are calculated at
 bare couplings (left to right) $\beta=5.2$ to~6.0 by 0.1, and then $\beta=7.0$ and~8.0.
Horizontal error bars are the size of the plotted symbols.
\label{fig:Beta}}
\end{figure}

Beginning with the scale factor $n=2$,
we thus calculate the running coupling $K/g^2$ for given bare coupling 
$\beta$ at the critical hopping parameter $\kappa=\kappa_c(\beta)$, first
 on a lattice with $4^4$ sites, which defines the scale $L=4a$, and then at
 the same $(\beta,\kappa)$ on an $8^4$ lattice, which gives the scale $2L$. 
We show the results in Fig.~\ref{fig:Beta}.  At large $\beta$ (corresponding 
to the perturbative regime, large $u$) the DBF agrees with the one-loop 
result $B(u,2)=-[Kb_1/(16\pi^2)]2\log2\simeq-1.43$; the dashed curve shows the two-loop result.
It is plain that the DBF departs from two-loop perturbation theory and crosses 
the axis in the neighborhood of $K/g^2=19$, or $g^2\simeq2.0$.
The corresponding bare coupling is $\beta\simeq 5.6$. 

For an indication of the dependence of the result on the lattice spacing
$a=L/4$, we have carried out a parallel calculation of $B(u,4/3)$ with
$a=L/6$.  This means calculating $K/g^2$ on a lattice with $6^4$ sites, at
the same $\beta$ values as above,%
\footnote{
We use the same values of $\kappa=\kappa_c$ and $u_0$ as above.}
 and comparing to the results on $8^4$
sites.  The first lattice defines the scale $L$ while the second now gives the scale $4L/3$.  This is economical because the same $8^4$ data are used in this calculation as in the preceding.

\begin{figure}[htb]
\vskip\baselineskip
\includegraphics*[width=\columnwidth]
{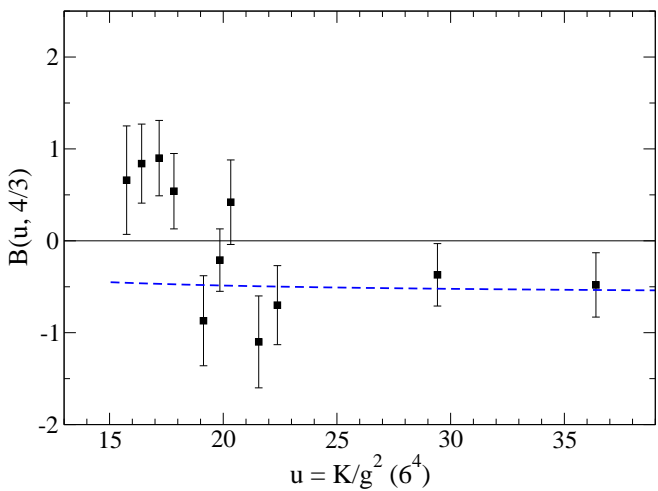}
\caption{Discrete beta function for the scale transformation $L\to4L/3$.
The lattice spacing is fixed such that $L=6a$.  
The dashed curve is the two-loop result. The data points are calculated at
the same bare couplings as in Fig.~\ref{fig:Beta}.
Horizontal error bars are the size of the plotted symbols.
\label{fig:Beta2}}
\end{figure}

These results are shown in Fig.~\ref{fig:Beta2} along with the two-loop prediction.
The latter is smaller than that for Fig.~\ref{fig:Beta} because the scale factor
is $4/3$ rather than 2.
In the lattice data, the crossing of zero (again near $g^2=2.0$) is evident.
This comparison of course does not take the place of a real scaling study.

\section{Nonconfinement and chiral symmetry}

We have also carried out calculations on lattices of volume $8^3\times N_t$
for $N_t=8$ and~12 with ordinary \hbox{(anti-)periodic} boundary conditions, at several values 
of $\kappa < \kappa_c$ in the same range of $\beta$.
These calculations show that $\beta=5.6$, with $\kappa=\kappa_c$, 
lies well on the weak-coupling side of the finite-volume crossover associated with confinement physics: The average
 Polyakov loop is large and the static $q\bar q$ potential
is entirely Coulombic in the range allowed by the volume.
This means that the lattices
of size $L=8$ are not large enough to contain the physics of confinement. 
As discussed above, this is essential for interpreting $g(L)$ as an 
effective coupling that characterizes the theory.

By QCD standards, a coupling $g^2/4\pi\simeq0.16$ is perturbative. This is in line with the lack of confinement in our $8^4$ lattices.
Nevertheless, this coupling may not be weak when considering
the possibility of spontaneous chiral symmetry breaking
for higher-representation fermions. Indeed, a quenched
study with color-sextet fermions found that the finite-temperature chiral transition occurs at $\beta \simeq 7.8$ with $N_t=4$, a far larger $\beta$ value than that of the deconfinement transition 
\cite{Kogut:1984sb}.
With our dynamical fermions, however, things are  different.
On lattices with $8^2\times12\times N_t$ sites, with $N_t=8$, we find the chiral
restoration crossover at $\kappa$ close to, but definitely below $\kappa_c(\beta)$.   Thus the coupling at the zero of the DBF 
doesn't break chiral symmetry either.


\section{DISCUSSION}

The simplest explanation of the zero of the DBF is that the strictly massless theory with two color-sextet fermions has an infrared-attractive fixed point,
leading to conformal physics without chiral symmetry breaking and without
confinement.  Accepting this conclusion, a major goal of further lattice
studies would be to understand the behavior of the same theory at
nonzero fermion mass.  In a truly conformal theory, the introduction of a fermion mass provides the only scale; that scale can sometimes play the role of 
an effective ultraviolet cutoff, and sometimes the role of an infrared cutoff.
This is an unfamiliar territory that contains many interesting questions.

Our results clearly do not allow for a continuum extrapolation.
More lattice volumes are needed, and it is also
desirable to compute the DBF for more scale ratios.
Such detailed information would allow the reconstruction of a 
continuous $\beta$ function.  
Appelquist, Fleming, and Neil \cite{Appelquist:2007hu} have recently presented 
an extensive SF analysis of the SU(3) gauge theory with $N_f=8$ and~12 flavors 
of fundamental fermions.  Their results are based on larger statistics as well
as a detailed continuum extrapolation. This allows them to conclude 
with certainty that the $N_f=12$ theory has an infrared fixed point. 

Our results for the DBF of the color-sextet theory do not preclude
more elaborate scenarios whereby the dynamics 
generates new effective degrees of freedom at some nonperturbative scale.  
This would typically lead to new relevant and/or marginal couplings; 
our single-parameter DBF would result from projecting the multiparameter
RG flow into a one-dimensional subspace. 

A concrete scenario in this direction was proposed by one of us in 
Ref.~\cite{Shamir:1989aj}.  One supposes that chiral symmetry breaks
spontaneously when the interaction is still Coulombic.
This allows for the existence of colored excitations.
The fermion condensate is then polarized in response
to an applied color field.  
As we move down in energy scale across the chiral transition,
the effective gauge coupling is screened by the collective response
of the fermion condensate.  This effect, if found, would be
the relativistic counterpart of the familiar dielectric polarization.  The scenario is described by an 
effective Lagrangian that indeed contains new couplings \cite{Shamir:1989aj}.
For a short while, the running of the gauge coupling {\em reverses its direction\/} from that
of an asymptotically free theory; a discrete sampling of this running 
can result in a zero of the DBF, imitating a fixed point.

Whatever the scenario, our DBF flatly contradicts continuum perturbative 
estimates. If confirmed by a true scaling study,
it would show that this theory is not like QCD with a small number of fundamental fermions, where the beta function is always negative;
and it is not like theories where the fermions condense and decouple
from the gauge fields, since in that case the beta function would behave as if $N_f=0$, becoming even more negative
than that of the fully coupled theory.
The simplest explanation of our DBF
is still an infrared-attractive fixed point, at an unexpectedly weak coupling. 
This would place the massless theory squarely in the conformal window:  It would prevent 
confinement and the spontaneous breaking of chiral symmetry, and remove
the theory from the list of candidates for walking.

We will present a detailed study of the physics of this model, so different from
 QCD, in a forthcoming paper.

\begin{acknowledgments}
We thank F.~Sannino and R.~Hoffmann for discussions.
This work was supported in part by the US Department of Energy and by the Israel Science Foundation under grant
no.~173/05.  Our computer code is based on version 7 of the publicly available code of the MILC collaboration~\cite{MILC}. 
\end{acknowledgments}

\end{document}